\documentclass[12pt]{article}
\usepackage{graphicx}
\usepackage{amssymb}
\usepackage{bm}
\evensidemargin \oddsidemargin
\def\1{{\bf 1}}
\def\0{{\bf 0}}

\def \F {{\cal F}}
\def \xim {\xi_{{\scriptscriptstyle -}}}

\def\b#1{{\mathbb #1}}
\def\nn{\nonumber \\}

\newcommand{\by}{{\bm y}}
\newcommand{\bY}{{\bm Y}}
\newcommand{\bx}{{\bm x}}
\newcommand{\bX}{{\bm X}}
\newcommand{\bu}{{\bm u}}

\newcommand{\bv}{{\bm v}}
\newcommand{\Ba}{{\bm \alpha}}
\newcommand{\bb}{{\bm \beta}}

\newcommand{\tbb}{\widetilde{\bm \beta}}

\newcommand{\bj}{{\bm j}}
\newcommand{\Be}{{\bm \epsilon}}
\newcommand{\bE}{{\bm E}}
\newcommand{\bB}{{\bm B}}
\newcommand{\bA}{{\bm A}}

\newcommand{\bee}{{\bm e}}
\newcommand{\Bp}{{\bm p}}

\newcommand{\be}{\begin{equation}}
\newcommand{\ee}{\end{equation}}
\newcommand{\bea}{\begin{eqnarray}}
\newcommand{\eea}{\end{eqnarray}}
\newcommand{\ba}{\begin{array}}
\newcommand{\ea}{\end{array}}
\newtheorem{prop}{Proposition}[section]
\newtheorem{naming}{Def.}[section]   %definition
\newcommand{\bdefi}{\medskip\begin{naming} ~ \it}
\newcommand{\edefi}{\end{naming} }% \bigskip}
\newcommand{\bp}{\begin{proof}}
\newcommand{\ep}{\end{proof}\par\vspace{10pt}\noindent}
\begin{document}

\title{On plane waves in diluted relativistic cold plasmas}
%\subtitle{Do you have a subtitle?\\ If so, write it here}

%\titlerunning{Short form of title}        % if too long for running head

\author{   Gaetano Fiore  \\  \\
        %  \and
Dip. di Matematica e Applicazioni, Universit\`a ``Federico II''\\
   V. Claudio 21, 80125 Napoli, Italy;\\         %\and
 I.N.F.N., Sez. di Napoli,
        Complesso MSA, V. Cintia, 80126 Napoli, Italy}

\date{}

\maketitle

\begin{abstract}

We briefly report on some exact results \cite{Fio13} regarding plane waves in a relativistic cold plasma. 
If the plasma, initially at rest, is reached by a  transverse plane electromagnetic 
travelling-wave, then  its motion has a very simple dependence
on this wave in the limit of zero density, otherwise can be determined
by an iterative procedure whose accuracy decreases with time or the plasma 
density. Thus one can describe in particular the impact of a very intense and short laser 
pulse onto a plasma and determine conditions for the  {\it slingshot effect} \cite{FioFedDeA13} to occur.  
The motion in vacuum of a charged test particle
subject to a wave of the same kind is also determined,  for any initial velocity. 

%Include keywords, PACS and mathematical
%subject classification numbers as needed.
%\keywords{Nonlinear PDEs, \and Laser-plasma interactions \and Laser-driven acceleration}
 %\subclass{76Wxx \and  82D10 \and 35Qxx \and  45Gxx}
\end{abstract}

\section{Introduction}
\label{intro}

The amazing developments of laser technologies today allow the
production of very intense (hundreds of TeraWatts), coherent electromagnetic (EM) waves
concentrated in very short pulses (tens of femtoseconds).
The interaction of such laser pulses with isolated electric charges or with continuous
matter is characterized by so fast, huge and highly nonlinear effects
that  traditional approximation schemes are seriously challenged. Even if the initial state of matter 
is not of plasma type, the huge kinetic energy $\kappa$ transfered to the electrons  almost immediately
ionizes matter locally into a plasma\footnote{Each level of ionization
(first, second,...) is practically complete if  the associated Keldysh parameter \
 $\Gamma_i\!:=\!\sqrt{U_i/\kappa}$ \  ($U_i$ is the associated ionization potential)
 fulfills \ $\Gamma_i\!\ll\!1$ \cite{Puk02,JovFedTanDeNGiz12}.}
(thereafter quantum effects are completely negligible).
The kinetic energies transfered to electrons and ions are also
 many orders of magnitude above the typical values of the thermal spectrum
 (even if the temperature is millions of ${}^{\circ}$K!),
therefore classical relativistic Magneto-Fluid-Dynamics (MFD) at zero temperature, with
its full nonlinearity, is a perfectly accurate framework while the pulse is passing, 
and also afterwards as long as dissipation
has not produced significant effects.

Due to the extremely high number of electrons in a plasma,  
moderate displacements w.r.t. 
ions generate huge electric fields that may in turn lead to extreme acceleration
of charged particles.
Understanding the underlying collective effect mechanisms would be crucial for many
scopes, from sheding light on some violent astrophysical phenomena
to conceiving a completely new kind of particle accelerators.
Today  accelerators are used in particular  for:

\begin{enumerate}

\item nuclear medicine, cancer therapy (PET, electron/proton therapy,...);
\item research in structural biology;
\item research in materials science;
\item food sterilization;
\item research in nuclear fusion (inertial fusion);
\item transmutation of nuclear wastes;
\item research in high-energy particle physics.

\end{enumerate}

Past and present-day acceleration technology (cyclotrons, synchrotrons, etc) 
relies on the interaction of radio-frequency (RF) EM waves   with `few' charged 
particles (those one wishes to accelerate)   over long distances. It has been developed for  purpose 7, 
%high-energy particle physics.
but has more recently found very important applications also for the other ones.
By now it is close to its structural limits.
The present or recent most powerful accelerators (LHC  and its predecessor LEP at CERN, 
SLAC in Stanford, Tevatron at Fermilab), which accelerate(d) particles up to energies of 100 - 1000 GeV, are (or were) already very big and expensive. 
In 2011 Tevatron closed because of budget cuts; building higher energy  accelerators would be prohibitive.
Much lower energies (100 - 300 MeV) are needed for other uses;  
but the still too large machines and high costs prevent the use on a large scale.
For instance, the CNAO center in Pavia - one of the few 
centers  for cancer treatment by hadron therapy - is based
on a 25-meters diameter synchrotron which has costed about 100 million Euro.

A lot of theoretical  efforts are being
made to conceive and construct new types of `table-top'
plasma-based acceleration machines, at least for energies of the order 
$100\div 1000$ MeV. A  beam of electrons
of about 200 MeV with very little energetic and angular spread has been
produced within a distance of few mm through the socalled Laser Wake Field (LWF) 
mechanism \cite{TajDaw79} in the bubble-regime \cite{FauEtAl04}
in experiments at the {\it \'Ecole Polytechnique} \cite{MalEtAl05}. 
The involved accelerations are thousands of times the ones
generated by RF-based accelerators.
  Present theoretical research on the subject is dominated by 
%the development and application of
numerical resolution programs of the MFD equations 
(`particle-in-cell' simulations, etc.) or their substitution 
by  (sometimes over-) simplified models. This leads to
a qualitative understanding of some phenomena, at best.
Susbtantial progress in a rigorous analytical study 
of the MFD equations would be highly welcome.

Here we briefly report about recent exact results \cite{Fio13}
applying to the differential (sect. \ref{plane})  and equivalent integral
equations (sect. \ref{integral}) ruling a relativistic cold plasma after the plane-wave Ansatz.  
If the plasma,  initially at rest, is reached by a   transverse plane EM 
travelling-wave, then the solution has a very simple dependence on 
the EM potential  in the limit of zero density  (sect. \ref{0densitysol}); 
otherwise the zero-density solution is a good approximation of the real
one as long as the back-reaction of the charges on the EM field 
can be neglected (i.e. for a time lapse decreasing with the plasma density), and can be 
corrected into better and better ones by an iterative procedure.
In sect. \ref{constdensity} we sketch how to use these results
to describe the impact of an ultra-intense and ultrashort laser pulse with a plasma
and determine conditions under which a new phenomenon named  {\it slingshot effect}  
\cite{FioFedDeA13}
%(i.e. the expulsion of energetic electrons in direction opposite to the laser pulse) 
should occur. The general motion of a
charged test particle in the above EM wave is  determined in sect. 
\ref{arbincond}. 

We first fix the notation and recall the basic equations.
We denote as $x=(x^\mu)=(x^0,\bx)=(ct,\bx)$ the spacetime coordinates
($c$ is the light velocity), $(\partial_\mu)\equiv(\partial/\partial x^\mu)=(\partial_0,\nabla)$,
as $(A^\mu)=(A^0,\bA)$ the EM potential,
as $F^{\mu\nu}=\partial^\mu A^\nu-\partial^\nu A^\mu$ the EM field,
and consider a collisionless plasma composed by
$k\ge 2$ types of charged particles (electrons, ions).
For $h\!=\!1,...,k$ let $m_h,q_h$ be the rest mass
and charge of the $h$-th type of particle (as usual, $-e$ the charge of electrons), 
 $\bv_h(x)$, $n_h(x)$
respectively the 3-velocity and the density (number of particles per unit volume)
of the corresponding fluid element located in position $\bx$ at time $t$. 
It is convenient to use 
%formulate the equations in terms of 
dimensionless variables like 
$$
\ba{l}
 \bb_h\!:=\!\bv_h/c, \qquad\gamma_h\!:=\!1/\sqrt{1\!-\!\bb_h^2}\\[6pt]
\mbox{4-vector velocity:}\quad u_h=( u_h^\mu)=(u^0_h,\bu_h)\!:=\!(\gamma_h,\gamma_h \bb_h)
=\left(\frac {p^0_h}{m_hc^2},\frac {{\bf p}_h}{m_hc}\right)
\ea
$$
(then \
$ u_h^\mu u_{h\mu}\!=\!1$, \ $\gamma_h\!=\!u_h^0\!=\!\sqrt{1\!+\!\bu_h^2}$, \
$\bb_h\!=\!\bu_h/\gamma_h$), and 
% (indices are raised and lowered by the Minkowski metric $\eta_{\mu\nu}\!=\!\eta^{\mu\nu}$, 
% with $\eta^{00}\!=\!1$, $\eta^{11}\!=\!-1$, etc.). \
$$
\ba{l}
\mbox{4-vector current density:}\quad (j^\mu)=(j^0,\bj)=
\left(\sum\limits_{h=1}^k q_h n_h,\sum\limits_{h=1}^k q_h n_h\bb_h\right)
\ea
$$

The Eulerian  and  Lagrangian descriptions  
%$f_h(x^0\!,\bx)$, $\tilde f_h(x^0\!,\bX)$ 
of an observable %of the $h$-th fluid 
are related by
\be
\tilde f_h(x^0\!,\bX)=f_h\left[x^0\!,\bx_h\!\left(x^0,\!\bX\right)\right]\quad\Leftrightarrow\quad
f_h(x^0\!,\bx)=\tilde f_h\!\left[x^0,\!\bX_h(x^0\!,\bx)\right],
\ee
where $\bx_h(x^0\!,\!\bX)$ is the  position at  time $t$ of the $h$-fluid element initially located
in $\bX$,  one requires $\bx_h\!\in\! C^1(\mathbb{R}^4)$ and the inverse
$\bX_h(x^0\!,\cdot)\!:\!\bx\!\mapsto\!\bX$ of  $\bx_h(x^0\!,\cdot)\!:\!\bX\mapsto \bx$ \ to exist.
Conservation of the %number of 
particles of the $h$-th fluid reads
\be
\ba{l}
\tilde n_h\left\vert\frac {\partial \bx_h} {\partial \bX}\right\vert=
\widetilde{n_{h0}}(\bX)\qquad
\Leftrightarrow \qquad  n_h\left\vert\frac {\partial \bX_h}{\partial \bx} \right\vert^{-1} =n_{h0} \label{n_hg}
\ea
\ee
and implies the continuity equation
\be
\frac {d n_h}{dx^0}\!+\!n_h\nabla\!\cdot\!\bb_h=\partial_0n_h\!+\!\nabla\!\cdot\!(n_h\bb_h)
=0;                          \label{clh'}
\ee
here $\frac {d}{dx^0}\!\!:=\!\!\frac {d}{cdt}\!\!=\!\!\partial_0\!+\!
\beta^l_h\partial_l\!=\!\frac{u_h^\mu}{\gamma_h}\partial_\mu$ is the
 {\it material} derivative for the $h$-th fluid, rescaled by $c$.
In the CGS system Maxwell's equations
and the (Lorentz) equations of motion of the fluids in Lorentz-covariant formulation read
\bea
&&\Box A^\nu-\partial^\nu(\partial_\mu A^\mu)
=\partial_\mu F^{\mu\nu}=4\pi j^\nu,\label{Maxwell}\\[8pt]
&&-q_h u_{h\mu} F^{\mu\nu}=m_hc^2 u_{h\mu} \partial^\mu u_h^{\nu}              
    \label{hom'}
\eea
(Eulerian description);  (\ref{hom'})$_{\nu=0}$ follows also from 
contracting (\ref{hom'})$_{\nu=l}$  with $ u^l_h$, $l\!=\!1,2,3$. 
Dividing (\ref{hom'})$_{\nu=l}$  by $\gamma_h$  gives
the familiar 3-vector formulation of (\ref{hom'})
\be
\ba{l}
q_h\left(\bE +\frac{\bv_h}c \wedge \bB
\right)=\partial_t\Bp_h+\bv_h \cdot \nabla \Bp_h
=\frac {d\Bp_h}{dt}
\ea\label{hom}
\ee
in terms of the electric and magnetic fields $E^l=F^{l0}=-\partial_0A^l-\partial_l A^0$, $B^l=-\frac 12
\varepsilon^{lkn}F^{kn}=\varepsilon^{lkn}\partial_k A^n$.
Given the initial momenta $\widetilde{\Bp_{h0}}$ and densities $\widetilde{n_{h0}}$ in (\ref{n_hg}) 
 the unknowns are $A^\mu,\bx_h, \bu_h$  and  the  equations 
to be solved are (\ref{Maxwell}-\ref{hom'})  and 
\be
\partial_0 \bx_h( x^0\!,\bX)=\bb_h\!\left[x^0\!,\bx_h({x^0}\!,\bX)\right].
%,\qquad\qquad\bx_h(x^0,\bX)=\bX  
 \label{lageul}
\ee

\section{Lorentz-Maxwell equations for plane waves}
\label{plane}

We restrict our attention to solutions such that for all $h$:
\bea
&& A^\mu, n_h, \bu_h \qquad \mbox{{\bf  depend only on }} z\!\equiv\!
x^3\!,x^0 \qquad \mbox{(plane wave Ansatz)},     \label{pw}\\[8pt]
&&\!\!\ba{ll}
A^\mu(x^0\!,z)\!=\!0,\qquad \bu_h(x^0\!,z)\!=\!\0,&\qquad \mbox{if }\:\:x^0\!\le\! z,\\ [6pt]
\exists\:\:\widetilde{ n_{h0}}(z) \quad \mbox{such that }
\:\sum_{h=1}^k\!q_h\widetilde{ n_{h0}}
\!\equiv\!0, \quad n_h(x^0\!\!,z)\!=\!\widetilde{ n_{h0}}(z)&\qquad \mbox{if }\:\:x^0\!\le\! z.
\ea
  \qquad         \label{asyc}
\eea
Eq. (\ref{pw}-\ref{asyc})  entail a partial gauge-fixing, imply
$\bB\!=\!\bB^{{\scriptscriptstyle\perp}}\!\!=\!{\hat{\bm z}}\!\wedge\!\partial_z\bA\!^{{\scriptscriptstyle\perp}}$, $\bE^{{\scriptscriptstyle\perp}}\!=\!-\partial_0\bA\!^{{\scriptscriptstyle\perp}}$,
\bea
%\bB\!=\!\bB^{{\scriptscriptstyle\perp}}\!\!=\!{\hat{\bm z}}\!\wedge\!\partial_z\bA\!^{{\scriptscriptstyle\perp}},
%\quad\bE^{{\scriptscriptstyle\perp}}\!=\!-\partial_0\bA\!^{{\scriptscriptstyle\perp}}\!,\qquad  
\bE(x)\!=\!\bB(x)\!=\!\0,\qquad
\bx_h(x)=\bx \qquad\quad \mbox{if }\:x^0\!\le\! z ,
 \label{conseq'}
\eea
$-\bA\!^{{\scriptscriptstyle\perp}}(x^0\!,z)\!=\!\!\int^{x^0}_{z}\!\!\!\!d\eta 
\bE^{{\scriptscriptstyle\perp}}(\eta,z)\!=\!\!\int^{x^0}_{-\!\infty }\!\!\!\!d\eta
\bE^{{\scriptscriptstyle\perp}}(\eta,z)$, so that 
$\bA\!^{{\scriptscriptstyle\perp}}$ becomes a {\it physical observable},
and the existence of the limits $n_h(-\infty,Z)\!=\!\widetilde{ n_{h0}}(Z)$, \
$\bx_h(-\infty,\bX)\!=\!\bX$. \ \ Hence we can adopt $-\infty$  as the
 `initial' time in the Lagrangian description.
The map $\bx_h(x^0\!,\cdot)\!:\!\bX\!\mapsto\! \bx$ is invertible iff
$z_h( x^0\!,Z)$ is strictly increasing  w.r.t.
$ Z\!\equiv\! X^3$ for each fixed $x^0$.
We shall abbreviate $Z_h\!\equiv\! X^3_h$.
Eq. (\ref{n_hg}) %, (\ref{clh'}), (\ref{djac}), (\ref{dtxx0''}),  (\ref{lageul}) respectively
becomes
\bea
&& \tilde n_h (x^0,Z) \partial_Z z_h( x^0,Z) =\widetilde{n_{h0}}( Z ),
 %(x^0,z)
\qquad \Leftrightarrow \qquad
n_h\, = \, n_{h0}\, \partial_z  Z_h.    \label{n_h}
%\\[6pt] &&\partial_0\left(\tilde n_h \frac{\partial z_h}{\partial  Z }\right)=0 ,\qquad\qquad
% \Leftrightarrow\qquad\qquad\frac {d n_h}{dx^0}=-n_h\partial_z\beta_h^z, \label{clh''}\\
%&& -\frac {d }{dx^0}\log \left[\partial_z  Z_h\right]=\partial_z\beta^z_h,            \label{djac'}\\
%&& \partial_0  Z =0\qquad\qquad\qquad \Leftrightarrow \qquad\qquad\qquad
%\frac {d  Z_h}{dx^0}= \partial_0Z_h+\beta_h^z \partial_z  Z_h=0,\label{dtzzx0''}
% \\ && \partial_0 z_h( Z , {x^0})=\tilde\beta_h^z[ Z ,  {x^0}]
% =\beta_h^z[{z}_h( Z , {x^0}), {x^0}],\qquad\qquad
% z_h( Z ,  X^0 )= Z .        \label{cauchyz}
%z_h( Z , {x^0})- Z
%=\int\limits^{x^0}_{\widetilde {x^0}}\! d{x^0}'\,\beta_h^z[{z}_h( Z , {x^0}'), {x^0}'].
%=\int\limits^{x^0}_{\tilde {x^0}}d{x^0}'\frac{u^z[{z}( Z , {x^0}'), {x^0}']}
%{\gamma_h[{z}( Z , {x^0}'), {x^0}']},
%\label{tzz}
\eea
$\partial_0  Z \!=\!0$ in the Eulerian description gives 
$\frac {d  Z_h}{dx^0}\!=\! \partial_0Z_h\!\!+\!\!\beta_h^z \partial_z  Z_h\!=\!0$ \ and by  (\ref{n_h}) 
\be
n_{h0} \,\partial_0  Z_h\!+\!n_h\beta^z_h=0.                        \label{j_h}
\ee

As known, eq. (\ref{hom'})$_{\nu=x,y}$ amounts to \ $\frac {d}{dx^0}(m_hc^2
\bu_h^{{\scriptscriptstyle\perp}}\!+\!q_h\bA\!^{{\scriptscriptstyle\perp}})\!=\!0$, \
which  implies \ $m_hc^2\tilde{\bu}_h^{{\scriptscriptstyle\perp}}\!+\!
q_h\tilde{\bA}\!^{{\scriptscriptstyle\perp}}\!=\!C(\bX)$; \ by  (\ref{asyc})  \ $C(\bX)\!\equiv\! 0$, whence 
\be
\tilde{\bu}_h^{{\scriptscriptstyle\perp}}\!= \frac {-q_h}{m_hc^2}\tilde{\bA}\!^{{\scriptscriptstyle\perp}}\qquad
\Leftrightarrow\qquad\bu_h^{{\scriptscriptstyle\perp}}\!= \frac {-q_h}{m_hc^2}\bA\!^{{\scriptscriptstyle\perp}},
 \label{hom'12}
\ee
which explicitly gives
$\bu_h^{{\scriptscriptstyle\perp}}$ in terms of $\bA\!^{{\scriptscriptstyle\perp}}$.
\ Eq. (\ref{Maxwell}) and the remaining (\ref{hom'}) become
\bea
&& (\ref{Maxwell})_{\nu=0}:\qquad\partial_z E^z=4\pi \sum_{h=1}^kq_hn_h
,\label{Maxwell0}\\%[8pt]
&& (\ref{Maxwell})_{\nu=z}:\qquad \partial_0E^z=-4\pi \sum_{h=1}^kq_hn_h\beta^z_h,
\label{Maxwell3}\\%[8pt]
&& (\ref{Maxwell})_{\nu=x,y}:\quad\:
\left[\partial_0^2\!-\!\partial_z^2\right]\!\bA\!^{{\scriptscriptstyle\perp}}=
\underbrace{ 4\pi
\sum\limits_{h=1}^kq_hn_h\bb^{{\scriptscriptstyle\perp}}_h }_{-
\frac{4\pi }{c^2}\bA\!^{{\scriptscriptstyle\perp}}\sum_{h=1}^k
\frac{q_h^2n_h}{m_h\gamma_h}  }\!\! ,
\qquad\qquad \label{Maxwell12}\\[4pt]
&& (\ref{hom'})_{\nu=0}:\qquad  
\frac{d\gamma_h}{dx^0}-\frac{q_hu^z_h E^z}{\gamma_h m_hc^2}-
\frac{q_h^2\partial_0(\bA\!^{{\scriptscriptstyle\perp}})^2}{2\gamma_hm_h^2c^4}=0\label{hom'0}\\[4pt]
&& (\ref{hom'})_{\nu=z}:\qquad  
\frac{du^z_h}{dx^0}-\frac{q_h E^z}{m_hc^2}+
\frac{q_h^2\partial_z(\bA\!^{{\scriptscriptstyle\perp}})^2}{2\gamma_hm_h^2c^4}=0\label{hom'3}
% && (\ref{hom'})_{\nu=z}\quad \Leftrightarrow\quad (\ref{hom})_{\nu=z}\!:
% \quad m_hc^2\frac{du^z_h}{dx^0}=q_h(E^z\!+\!\bb_h^{{\scriptscriptstyle\perp}}\!\cdot\!\partial_z\bA\!^{{\scriptscriptstyle\perp}})
% \stackrel{(\ref{hom'12})}{=}q_hE^z\!-\!\frac{m_hc^2}{\gamma_h}
% \bu_h^{{\scriptscriptstyle\perp}}\!\cdot\!\partial_z\bu_h^{{\scriptscriptstyle\perp}} \nn[8pt]
% && \qquad \qquad \Leftrightarrow\qquad\qquad
% \frac{q_h} {m_hc^2}E^z=\frac{du^z_h}{dx^0}+\frac 1{2\gamma_h}
% \partial_z(\bu_h^{{\scriptscriptstyle\perp}}{}^2), \label{hom'3}\\[8pt]
\eea
The independent unknowns in (\ref{Maxwell0}-\ref{hom'3}) are $\bA\!^{{\scriptscriptstyle\perp}}, u_h^z,E^z$
(all observables).
%We neither need nor care to determine  $A^0,A^z$ such that
%$ E^z\!=\!-\partial_0A^z\!-\!\partial_z A^0$ by completing the gauge-fixing.

\subsection{Magnetic and ponderomotive force}
\label{0densitysol}

The term \ $F\!\!_{hm}^{\,\, z}\!:=\!-\partial_zq_h^2\bA\!^{{\scriptscriptstyle\perp}2}/2\gamma_hm_hc^2$ 
\ in (\ref{hom'3}) is the longitudinal magnetic part 
$q_h(\bb_h\! \wedge\! \bB)^z$ of the Lorentz force [cf. (\ref{hom})]; 
\ $U_{hm}\!:=\!-q_h^2\bA\!^{{\scriptscriptstyle\perp}2}/2m_hc^2$
acts like a `time-dependent potential energy'. For fixed $x^0$ let $\bar z\!<\!x^0$  the right extreme of 
 supp$\bA\!^{{\scriptscriptstyle\perp}}$: then $\bA\!^{{\scriptscriptstyle\perp}2}(x^0\!,z)\!=\!0$ for $z\!\ge\!\bar z$,
whereas  $\bA\!^{{\scriptscriptstyle\perp}2}(x^0\!,z)$ is positive and necessarily strictly decreasing for 
$z$ in a suitable interval $[z',\bar z[$. Then $F\!\!_{hm}^{\,\, z}$ acts as a positive longitudinal force on {\it all} the electric charges located at $z\!\in[z',\bar z[$.

\begin{figure}[ht]
%\begin{center}
\includegraphics[width=5.5cm]{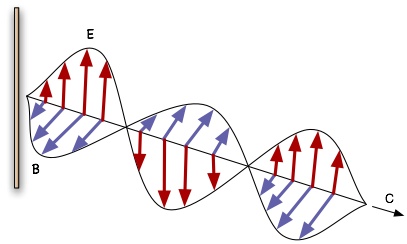}
\hfill
\includegraphics[width=6cm]{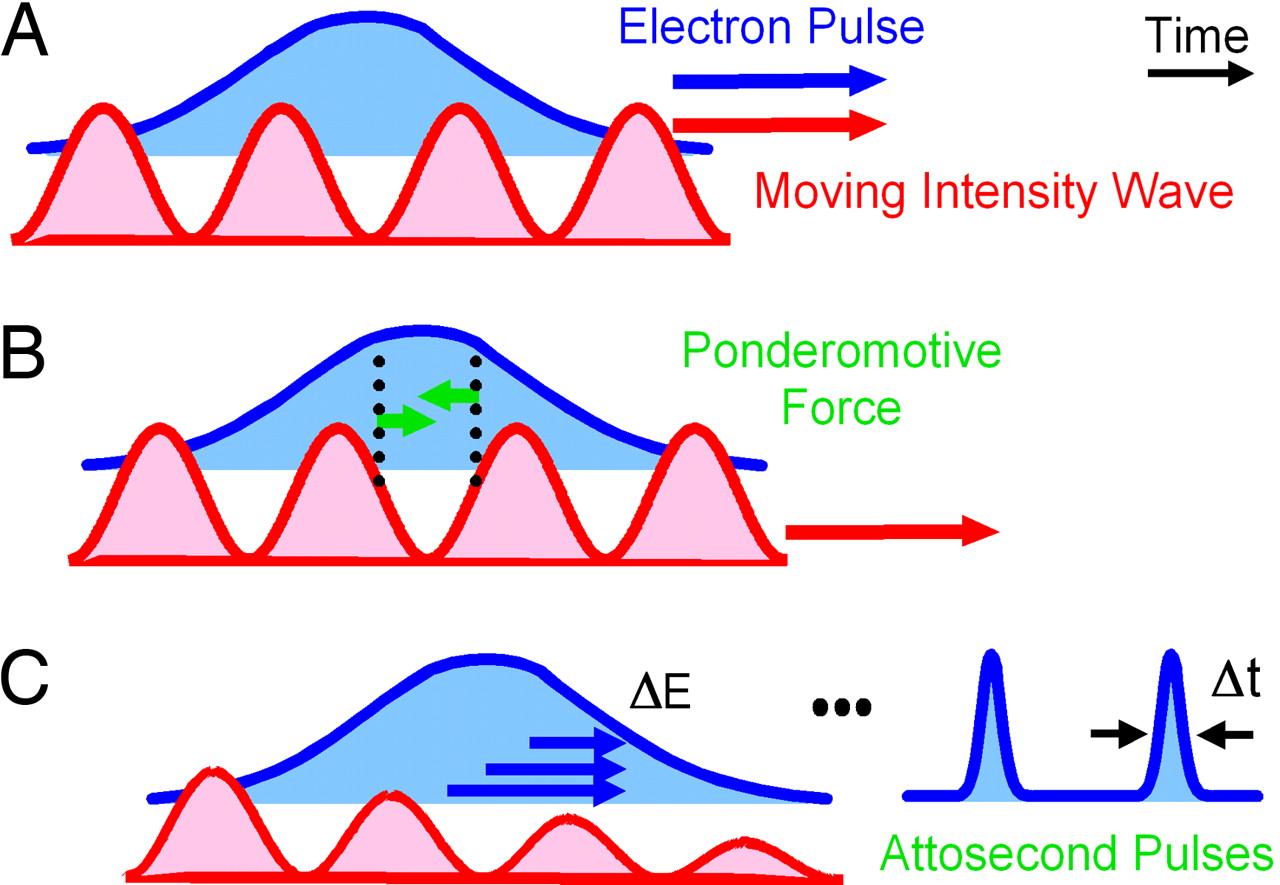}
%\end{center}
\caption{Schematic plots of a linearly polarized transverse wave (left) and of its interaction 
with a density wave of  electrons (right).}
\label{F2large}       
\end{figure}

In the prototypical cases  of a modulated monochromatic transverse wave
\be
\ba{ll}
\qquad\qquad\bE^{{\scriptscriptstyle \perp}}\!(x^0\!,z)\!=\!\Be^{{\scriptscriptstyle \perp}}\!(x^0\!\!-\!\!z),\qquad
&\qquad\Be^{{\scriptscriptstyle \perp}}(\xi)\!=\!\epsilon_s(\xi)\Be_o^{{\scriptscriptstyle \perp}}\!(\xi),\\[8pt]
\Be_o^{{\scriptscriptstyle \perp}}\!(\xi)\!=\! {\hat\bx}\cos k\xi,      
\quad\qquad\qquad  \Be_p^{{\scriptscriptstyle \perp}}\!(\xi)\!=\! 
{\hat\bx}\sin k\xi          &\qquad \mbox{(linearly polarized), or}\\[8pt]
\Be_o^{{\scriptscriptstyle \perp}}\!(\xi)\!=\!{\hat\bx}\cos k\xi\!+\!{\hat\by}\sin k\xi, 
\quad  \Be_p^{{\scriptscriptstyle \perp}}\!(\xi)\!=\! -\frac 1k\Be_o^{{\scriptscriptstyle \perp}\prime}
%{\hat\bx}\sin k\xi\!-\!{\hat\by}\cos k\xi 
&\qquad \mbox{(circularly polarized),}
\ea                                                                           \label{prototype}
\ee
with amplitude not varying significantly over $\lambda\!:=\!2\pi\!/\!k$, e.g. 
$\lambda |\epsilon_s'\!(\!\xi\!)\!/\!\epsilon_s\!(\!\xi\!)|\!\le\!\delta\!\ll\!1$ for all $\xi$,
then \ $\bA\!^{{\scriptscriptstyle\perp}}\!(x^0\!,z)\!=\! 
\left\{\frac{1}{k}\epsilon_s\left[\Be_p^{{\scriptscriptstyle \perp}}\!+\!O(\delta)\right]\right\}\!(x^0\!\!-\!\!z)$. \
The {\it ponderomotive force} \ $F\!\!_{hp}^{\,\, z}\!:=\! \langle F\!\!_{hm}^{\,\, z}\rangle$ \
 ($ \langle \,\,\rangle$ stands for the average over a period $\lambda$)
plays a crucial role in the LWF acceleration and in the slingshot effect. Up to $O(\delta)$ one finds
\be
\ba{ll}
F\!\!_{hm}^{\,\, z}\!=\!\left[\mu_h(\epsilon_s\Be_p^{{\scriptscriptstyle \perp}}\!)^2{}'\right]\!(x^0\!\!-\!\!z),
\quad F\!\!_{hp}^{\,\, z}\!=\!\frac 12\!\left[\mu_h(\epsilon_s^2){}'\right]\!(x^0\!\!-\!\!z)
%\!+\!O\big[\big(\frac\lambda l\big)^3\big]
\qquad &\mbox{lin. polarized,}\\[10pt]
F\!\!_{hm}^{\,\, z}\!=\!\left[\mu_h(\epsilon_s\Be_p^{{\scriptscriptstyle \perp}}\!)^2{}'\right]\!(x^0\!\!-\!\!z)
\!=\!\left[\mu_h(\epsilon_s^2){}'\right]\!(x^0\!\!-\!\!z) \equiv F\!\!_{hp}^{\,\, z} 
\qquad &\mbox{circ. polarized;}
\ea\ee
here %$\omega\!=\!2\pi\nu\!=\!2\pi c/\lambda$ and 
we abbreviated $\mu_h\!:=\!\lambda^2 q_h^2/8\pi^2\gamma_hm_hc^2$,
%$\mu_h\!:=\!q_h^2/2\omega^2\gamma_hm_h$. 
and we have used that $\Be_p^{{\scriptscriptstyle \perp}2}\!=\!1$ for circular polarization.
  Hence the ponderomotive force \ $F\!\!_{hp}^{\,\, z}(x^0\!,z)$ \
is  positive (resp. negative) for {\it all} $h$ if $\epsilon_s^2(\xi)$ 
is increasing (resp. decreasing) at $\xi\!:=\!x^0\!\!-\!\!z$ (fig. \ref{F2large} - right A). 
Therefore, while the transversal motion is oscillatory
with period $\lambda$ and averages to zero, the longitudinal motion is 
ruled in average by $\epsilon_s$ on the much larger scale $l$;
in the case of a linearly polarized wave
 the rapid spatial oscillations of $\Be_p^{{\scriptscriptstyle \perp}2}$ have 
the additional effect of modulating the densities, especially of the electrons (the lightest particles),
into equi-spatiated bunches (see fig. \ref{F2large} - right B,C).
Moreover, for large amplitudes \ ($u_h\gg 1$) \  the  direction of $\bv_h$ is 
close to the longitudinal one for most of the time.

\section{The zero-density solutions}
\label{0densitysol}

\begin{prop} \cite{Fio13} \
If $\Ba^{{\scriptscriptstyle \perp}}(\xi)\!\in\! C^2(\b{R},\b{R}^2)$ and
$\Ba^{{\scriptscriptstyle \perp}}(\xi)\!=\!0$ for \ $\xi\!\le \!0$ \ then
\be
\ba{lll}
\bA\!^{{\scriptscriptstyle\perp}}(x) \!=\! 
%\bA\!^{{\scriptscriptstyle\perp(0)}}\!(\xi)\!:=\! 
\Ba^{{\scriptscriptstyle \perp}}(\xi),\quad\:\: \xi\!\!:=\!x^0\!\!-\!z,\qquad &
n_h \!=\!n_{h}^{{\scriptscriptstyle (0)}}  \!:=\!0,\qquad \: &
E^z \!=\!E^{z{\scriptscriptstyle (0)}}\!\!:=\! 0,   \\[8pt]
\bu_h^{{\scriptscriptstyle\perp}}(x)\!=\!\bu_{h}^{{\scriptscriptstyle \perp(0)}}(\xi)\!:=\!
\frac {-q_h}{m_hc^2}\Ba^{{\scriptscriptstyle \perp}}(\xi),
\quad \: &
u_h^z\!=\!u_{h}^{{\scriptscriptstyle z(0)}}
\!\!:=\!\frac 12\bu_{h}^{{\scriptscriptstyle \perp(0)}}{}^2\!, \qquad \:
&  \gamma_h\!=\!\gamma_{h}^{{\scriptscriptstyle (0)}}
\!\!:=1+\!u_{h}^{{\scriptscriptstyle z(0)}},
 \ea \label{n=0'+}
\ee
which depend on $x$ only through $\xi$, solve
(\ref{hom'12}-\ref{hom'3}) and  (\ref{asyc}).
\label{prop1}
\end{prop}

\noindent Let $s_h\!:=\!\gamma_h\!-\! u^z_h$. The difference of eqs. (\ref{hom'0}-\ref{hom'3}) 
gives the equivalent equation
\be
 \frac {ds_h}{d{x^0}}=\frac {q_h^2}
{2m_h^2c^4\gamma_h} \left(\partial_0+ \partial_z\!\right)\bA\!^{{\scriptscriptstyle\perp}}{}^2
\!-\!\frac{s_h}{\gamma_h}\frac{q_h E^z}{m_hc^2}                   \label{bla'}
\ee
The  assumptions imply $\frac {ds_h}{d{x^0}}\!=\!0$, \ whence \ $s_h\!\equiv\!1$, \
which is the main step of the proof.
Eqs. (\ref{n=0'+}) give travelling-waves determined solely by the assigned
$\Ba^{{\scriptscriptstyle \perp}}$ and moving in the 
$\hat z$ direction  with phase velocity equal to $c$;
they make up a weak solution if $\Ba^{{\scriptscriptstyle \perp}}$
is less regular, e.g. $\Ba^{{\scriptscriptstyle \perp}}(\xi)\!\in\! C(\b{R},\b{R}^2)$
and  $\Ba^{{\scriptscriptstyle \perp\prime}}\!=\!\Be^{{\scriptscriptstyle \perp}}$ is continuous 
except in a finite number of points of  finite discontinuities (e.g. at the wavefront).
At no time any particle can move in the negative $z$-direction because $u_{h}^{{\scriptscriptstyle z(0)}},\beta_{h}^{{\scriptscriptstyle z(0)}}$ are nonnegative-definite; the latter are the result of the
acceleration by the $z$-component $F\!\!_{hm}^{\,\, z}$ of the magnetic force.

We introduce the following primitives of \ $\bu_h^{{\scriptscriptstyle (0)}},\gamma_h^{{\scriptscriptstyle (0)}}$:
\be
\bY\!_h(\xi)\!:=\!\int^\xi_0\!\!\!\! d\xi' \,\bu_h^{{\scriptscriptstyle (0)}}\!(\xi'),
%\!=\!\!\int^\xi_{-\infty}\!\!\!\!\!  d\xi' \bu_h^{{\scriptscriptstyle (0)}}\!(\xi'),
\qquad\quad\Xi_h(\xi)\!:=\!\int^\xi_0\!\!\!\!  d\xi'\, \gamma_h^{{\scriptscriptstyle (0)}}(\xi')\!=\!
\xi  \!+\! Y^3_h(\xi);                                  \label{defYXi}
\ee
As $u_{h}^{{\scriptscriptstyle z(0)}}\!\ge\! 0$,  $Y^3_h(\xi)$ is increasing, \
$\Xi_h(\xi)$ \ is strictly increasing and invertible.
 
\begin{prop} \cite{Fio13} \ Choosing $\bu_h\!\equiv\!\bu_h^{{\scriptscriptstyle (0)}}$,
the solution \ $\bx^{{\scriptscriptstyle (0)}}_{h}(x^0,\bX)$ \ of the ODE
(\ref{lageul}) with the initial condition \
$\bx_h(x^0,\bX)\!=\!\bX  {}$ \ for $x^0\!\le\! Z$,
%$\bx_h(x^0,\bX)\stackrel{x^0\!\to\!-\infty}{\longrightarrow}\bX  {}$, \
and (for fixed $x^0$) its inverse \ $\bX^{{\scriptscriptstyle (0)}}_{h}(\bx,x^0)$ \ are given by:
\be
\ba{l}
z^{{\scriptscriptstyle (0)}}_{h}\!(x^0\!,Z)=x^0\!-\!\Xi_h^{-1}\!
\left( x^0\!-\! Z \right),\\[6pt]
 Z^{{\scriptscriptstyle (0)}}_{h}\!(x^0\!,z)=x^0\!-\!\Xi_h\!
\left( x^0\!-\! z \right)=z\!-\! Y^3_h(x^0\!-\! z),\\[6pt]
\bx^{{\scriptscriptstyle \perp(0)}}_{h}(x^0,\bX )=\bX^{{\scriptscriptstyle \perp}}
\!+\!\bY^{{\scriptscriptstyle \perp}}_h\!\left[x^0\!-\!
z^{{\scriptscriptstyle (0)}}_{h}\!(x^0\!,Z )\right],\\[6pt]
\bX^{{\scriptscriptstyle \perp(0)}}_{h}(x^0\!,\bx)=
\bx^{{\scriptscriptstyle \perp}} \!-\!
\bY\!^{{\scriptscriptstyle \perp}}_h\!\left(x^0\!-\!z\right).
\ea \label{hatxtxp}
\ee
%\be
%\xi^{{\scriptscriptstyle (0)}}_{h}\!(\tilde \xi )=\Xi_h^{-1}\!\left(\tilde\xi\right),
%\qquad \tilde\xi^{{\scriptscriptstyle (0)}}_{h}\!(\xi)=\Xi_h(\xi).\label{xitxi}
%\ee
These functions fulfill 
\be
\partial_0  Z ^{{\scriptscriptstyle (0)}}_h \!=\!-u_h^{{\scriptscriptstyle  z(0)}}\!,
\quad\partial_z  Z ^{{\scriptscriptstyle (0)}}_h \!=\!\gamma_h^{{\scriptscriptstyle  (0)}}\!
,\quad\partial_Z z^{{\scriptscriptstyle (0)}}_h
\!=\!\frac1 {\widetilde{\gamma_h}^{{\scriptscriptstyle (0)}}}\!,\quad
 \partial_Z \bx^{{\scriptscriptstyle \perp(0)}}_h\!=\!- 
\tbb_{h}^{{\scriptscriptstyle \perp(0)}}\!.                    \label{hatdtzz'}
\ee
\label{prop2}
\end{prop}
From (\ref{hatxtxp}) it follows that  the longitudinal displacement
of the $h$-th type of particles w.r.t. their initial position $\bX$ at time $x^0$  is
\be
\Delta z_h^{{\scriptscriptstyle (0)}}\! ( x^0\!,\! Z ) :=
z_h^{{\scriptscriptstyle (0)}}\! ( x^0\!,\! Z )\!-\!Z\,=\, Y^3_h\!\left[\Xi_h^{-1}\!
\left( x^0\!-\! Z \right)\right].                 \label{displace}
\ee
By (\ref{n=0'+}) the evolution of $\bA^{{\scriptscriptstyle \perp}}$ amounts to a translation
of the graph of $\Ba^{{\scriptscriptstyle \perp}}$.
Its value $\check \Ba{{\scriptscriptstyle \perp}}\!:=\!\Ba{{\scriptscriptstyle \perp}}(\check \xi)$ 
at some point  $\check \xi$ 
%(e.g. a minimum or a maximum) 
reachs the particles initially located in $Z$ at the time $\check x_h^0(\check \xi,Z)$ such that 
\be
\check x_h^0\!-\!\check \xi=z_h^{{\scriptscriptstyle (0)}}\! (\check x^0\!,\! Z )
\!\stackrel{(\ref{hatxtxp})_1}{=}\!\check x_h^0\!-\!\Xi_h^{-1}\! \left[\check x_h^0\!-\! Z \right] 
\qquad\Leftrightarrow \qquad
\check x_h^0(\check \xi,Z)\!=\!\Xi_h(\check \xi)\!+\! Z,                        \label{chain}
\ee
in the position $z_h^{{\scriptscriptstyle (0)}} (\check x^0\!, Z )=
\Xi_h(\check \xi)\!+\! Z\!-\!\check \xi =Y^3_h(\check \xi)\!+\! Z $. The corresponding displacement 
of these particles is independent of $Z$ and equal to
\be
\zeta_h=
\Delta z_h^{{\scriptscriptstyle (0)}}\! \left[\check x_h^0 (\check \xi,Z), Z \right]=Y^3_h(\check \xi)
\label{checkzeta}
\ee

\section{Motion of test particles with arbitrary initial conditions}
\label{arbincond}

\begin{figure}[ht]
%\begin{center}
\includegraphics[width=6cm]{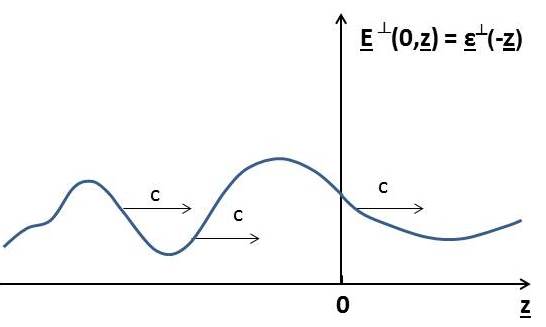}\hfill
\includegraphics[width=6cm]{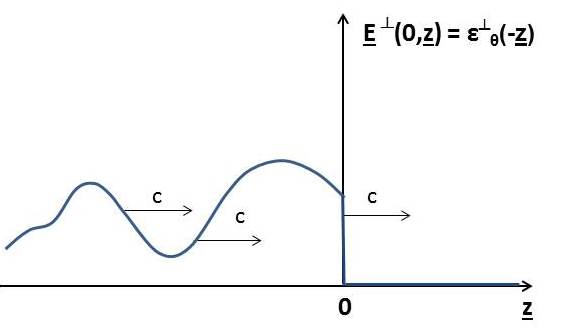}
%\end{center}
\caption{The original  $\underline\bE\!^{{\scriptscriptstyle\perp}}$ (left) and its 'cut' counterpart (right)
as functions of $\underline z$ at  $\underline x^0=0$.}
\label{CutEPlot}       
\end{figure}

Eq. (\ref{hatxtxp}) describes also the motion of a  {\it single} test particle of 
charge $q_h$ and mass $m_h$ starting from position $\bX$ with velocity 
$\0$ at sufficiently early time, i.e. before the EM wave arrives.
These results for a single test particle can be obtained also by solving 
the Hamilton-Jacobi equation \cite{LanLif62}.
Let now \ $\bE\!^{{\scriptscriptstyle\perp}}(x)\!=\!\Be\!^{{\scriptscriptstyle\perp}}( x^0\!-\! \bx\cdot\bee)$, \
$\bB\!^{{\scriptscriptstyle\perp}}\!=\!\bee\!\wedge\!\bE\!^{{\scriptscriptstyle\perp}}$ 
($\bee$ is the unit vector of the direction of propagation of the wave)
be an arbitrary free transverse plane EM travelling-wave 
(we {\it no longer} require $\bE\!^{{\scriptscriptstyle\perp}}, \bB\!^{{\scriptscriptstyle\perp}}$ to vanish for  $x^0\!-\! \bx\cdot\bee\!<\!0$).
The {\it general solution} $\bx_h(x^0)$ of the Cauchy problem (\ref{hom}-\ref{lageul}) 
with initial conditions \ $\bx_h(0)\!=\!\bx_0$, $\frac{d\bx_h}{dx^0}(0)\!=\!\bb_0$ \
under the action of such an EM travelling-wave can be now determined by redution  to the previous one as follows. 
One can do a Poincar\'e transformation $P=TRB$
to a new reference frame $\underline\F$ where the initial velocity and position are zero
 and the wave propagates in the positive $z$ direction: one first
finds the boost $B$ from the initial reference frame $\F$ to a new one $\F'$ where $\bb_0'\!=\!\0$
(this maps the transverse plane electromagnetic wave into a new one), then a rotation
$R$ to a reference frame $\F''$ where the 
plane wave propagates in the positive $z$-direction,
finally the translation $T$ to the reference frame $\underline\F$ where also $\underline\bx_0\!=\!\0$.
Naming $\underline x^\mu$ the spacetime coordinates and $\underline A^\mu,\underline F^{\mu\nu},...$ 
the fields w.r.t.   $\underline \F$, it is 
 $\frac{d\underline\bx_h}{d\underline x^0}(0)\!=\!\0$,  $\underline\bx_h(0)\!=\!\0$,
and $\underline\bE\!^{{\scriptscriptstyle\perp}}(x)\!=\!\underline\Be\!^{{\scriptscriptstyle\perp}}(\underline x^0{}\!-\!\underline z)$, \ 
$\underline\bB\!^{{\scriptscriptstyle\perp}}\!=\!\hat{{\underline{\bm z}}}\!\wedge\!\underline\bE\!^{{\scriptscriptstyle\perp}}$.
Since the part of the EM which is already at the right of the particle at  $\underline x^0\!=\!0$ will not 
come in contact with the particle nor affect its motion,
the solution $\underline x_h(\underline x^0)$ of the Cauchy problem w.r.t. 
$\underline\F$ does not change if we replace 
$\underline\Be^{{\scriptscriptstyle\perp}}\!(\underline x^0\!-\!\underline z)$ by the `cut' counterpart
$\Be_{{\scriptscriptstyle\theta}}^{{\scriptscriptstyle\perp}}\!(\underline x^0\!-\!\underline z)\!:=\!
\underline\Be^{{\scriptscriptstyle\perp}}\!(\underline x^0\!-\!\underline z)\,\theta(\underline x^0\!-\!\underline z)$
($\theta$ stands for the Heaviside step function), see fig. \ref{CutEPlot}. Clearly 
$\Be^{{\scriptscriptstyle\perp}}_{{\scriptscriptstyle\theta}}(\xi)$ and 
$\Ba\!^{{\scriptscriptstyle\perp}}_{{\,\scriptscriptstyle\theta}}(\xi)\!:=\!\!\int^\xi_0\!\! d\xi' \,\Be_{{\scriptscriptstyle\theta}}^{{\scriptscriptstyle\perp}}\!(\xi')$
fulfill $\Be_{{\scriptscriptstyle\theta}}^{{\scriptscriptstyle\perp}}(\xi)\!=\!\Ba\!_{{\,\scriptscriptstyle\theta}}^{{\scriptscriptstyle\perp}}(\xi)\!=\!\0$
if $\xi\!\le\!0$; therefore, denoting as $\bu^{{\scriptscriptstyle (0)}}_{h{\scriptscriptstyle\theta}}(\xi),\bx^{{\scriptscriptstyle (0)}}_{h{\scriptscriptstyle\theta}}(\underline x^0\!,\bX ),...$ the functions of the previous section obtained choosing 
 $\Ba\!^{{\scriptscriptstyle\perp}}(\xi)\equiv\Ba\!_{{\,\scriptscriptstyle\theta}}^{{\scriptscriptstyle\perp}}(\xi)$,  we find
 $\underline \bx_h(\underline x^0)\!=\!\bx^{{\scriptscriptstyle (0)}}_{h{\scriptscriptstyle\theta}}(\underline x^0\!,\0 )$.
The solution in $\F$ is finally obtained applying the inverse Poincar\'e transformation $P^{-1}$ 
to $\underline \bx_h(\underline x^0)$.

\section{Integral equations for plane waves}
\label{integral}

We reformulate the PDE's  as integral equations. Using (\ref{n_h}-\ref{j_h}) 
one proves  

\begin{prop} \cite{Fio13} \ For any $\bar Z\!\in\!\b{R}$ eq.
(\ref{Maxwell0}-\ref{Maxwell3}) and (\ref{asyc}) are solved by
\be
  E^{{\scriptscriptstyle z}}(x^0,z)=4\pi \sum\limits_{h=1}^kq_h
\widetilde{N}_h[ Z_h(x^0\!,z)], \qquad
\qquad\widetilde{N}_h(Z):=\int^{Z}_{\bar Z}\!\!\! d Z'\,\widetilde{n_{h0}}(Z');
\label{expl}
\ee
the neutrality condition (\ref{asyc})$_3$ implies $\sum\limits_{h=1}^kq_h \widetilde{N}_h(Z)\equiv 0$.
\end{prop}
Formula  (\ref{expl}) gives the solution of (\ref{Maxwell0}-\ref{Maxwell3})
explicitly in terms of the  initial densities, up to determination of the functions $ Z_h(x^0\!,z)$.

Using the Green function \ $G(x^0,z)\!=\!\frac 12 \theta(\xi) \theta(\xim)\!=\!\frac 12 \theta(x^0\!-\!|z|)$ \ of the d'Alembertian $\partial_0^2\!-\!\partial_z^2=4\partial_+\partial_-$ \ 
[$\theta$ is the Heaviside step function, \  $\xim\!\!:=\!x^0\!\!+\!z$, $\partial_-\!:=\!\partial/\partial_{\xim}\!=\!\frac 12(\partial_0\!+\!\partial_z)$], we can rewrite eq. (\ref{Maxwell12}) with initial
conditions at $x^0\!=\!X^0$  as the integral equation
\be
\ba{l}
\bA\!^{{\scriptscriptstyle\perp}}(x^0\!,z)-
\bA\!_{{\scriptscriptstyle f}}^{{\scriptscriptstyle\perp}}(x^0,z)=-\!\!
\displaystyle\int_{D\!^{X^0}_x}\!\!\!\!\!\! d^2x'\,
2\pi \sum\limits_{h=1}^kq_h\left[n_h\bb^{{\scriptscriptstyle\perp}}_h
\right]\!\!({x^0}'\!, z')         
\ea  \label{inteq1}
\ee
where $\bA\!_{{\scriptscriptstyle f}}^{{\scriptscriptstyle\perp}}(x^0\!,z)\!=\!\Ba\!^{{\scriptscriptstyle\perp}}\!(x^0\!\!-\!z)+\Ba\!_{{\scriptscriptstyle -}}^{{\scriptscriptstyle\perp}}\!(x^0\!\!+\!z)$
is determined by the initial conditions,
\bea
 \ba{l}
D\!^{X^0}_x\!\!:=\!\{ x'\:|\:  X^0 \!\!\le\! {x^0}'\! \!\le\! x^0\!,
\, |z\!-\!z'|\!\le\!{x^0}\!\!-\!{x^0}'\! \}=\{ x'\:|\: 2 X^0\! \!\le\!\xi'\!\!+\!\xim'\!,
\, \xi'\!\le\! \xi, \, \xim'\!\!\le\! \xim\}.   
\ea         \nonumber % \label{defs}
\eea
If beside (\ref{asyc}) {\bf we assume \ $n_{h}(0,z)\!=\!0$ for $z\!<\!0$}, \
  for $x^0\!\le\! 0$ the EM wave is free and 
$\bA\!^{{\scriptscriptstyle\perp}}$ is of the form
$\bA\!^{{\scriptscriptstyle\perp}}( x^0 ,z)\equiv \Ba\!^{{\scriptscriptstyle\perp}}( x^0 \!-\!z)$,
with $\Ba\!^{{\scriptscriptstyle\perp}}(\xi)\!=\!0$  for  $\xi\!\le\! 0$. 
%$Q\bA\!^{{\scriptscriptstyle\perp}}|_{x^0\!=\!  0 }\equiv 0$. 
Then in   (\ref{inteq1}) we may choose % the initial time 
$X^0\!=\!0$   [hence
$\widetilde{n_{h0}}(Z)\!\equiv\! n_{h}(0,\!Z)$] and set % $D\!_{x}\!=\!D\!_{0\!,x}$,
$\bA\!_{{\scriptscriptstyle f}}^{{\scriptscriptstyle\perp}}(x^0,z)
\!=\!\Ba\!^{{\scriptscriptstyle\perp}}( x^0 \!-\!z)$.

\noindent
In the Lagrangian description (\ref{bla'}) reads 
$\tilde\gamma_h\partial_0 \tilde s_h=\tilde s_h\tilde\varepsilon^z_h\!+\!
 \widetilde{\partial_-\bu_h^{{\scriptscriptstyle\perp}}{}^2}$;
the Cauchy problem with initial condition $\tilde s_{h0}\!\equiv\! 1$ is
equivalent to the integral equation
\be
\tilde s_h = e^{\int\limits^{x^0}_{ 0 }\!\!d\eta \,\tilde \mu_h\!(\eta, Z )}
\!\!\!+\!\!\displaystyle\int\limits^{x^0}_{ 0 }\!\!
d\eta\,e^{\int\limits^{x^0}_{\eta}\!\!d\eta'\tilde \mu_h\!(\eta', Z )} \left[\frac { \widetilde{\partial_-\bu_h^{{\scriptscriptstyle\perp}}{}^2}}{\tilde\gamma_h}\right]\! (\eta, Z ),
\qquad \tilde \mu_h\!:=\!\frac {-q_h \tilde E^z}{m_hc^2\tilde\gamma_h}. 
\label{inteq2}
\ee
$u_h^z,\gamma_h,\bb_h^{{\scriptscriptstyle\perp}},\beta_h^z$ can be recovered from $s_h,\bu_h^{{\scriptscriptstyle\perp}}$ through the formulae
\be
\ba{ll}
\displaystyle\gamma_h\!=\!\frac {1\!+\!\bu_h^{{\scriptscriptstyle\perp}}{}^2\!\!+\!s_h^2}{2s_h}, \qquad\qquad & \displaystyle\bb_h^{{\scriptscriptstyle\perp}}\!=\! \frac{\bu_h^{{\scriptscriptstyle\perp}}}{\gamma_h} \!=\!\frac{2s_h\bu_h^{{\scriptscriptstyle\perp}}}
 {1\!+\!\bu_h^{{\scriptscriptstyle\perp}2}\!\!+\!s_h^2 },\\[18pt]
\displaystyle u_h^z\!=\!\frac {1\!+\!\bu_h^{{\scriptscriptstyle\perp}}{}^2\!\!-\!s_h^2}{2s_h}, 
 \qquad\qquad & \displaystyle
\beta_h^z\!=\! \frac{u_h^z}{\gamma_h}\!=\!\frac{1\!+\!\bu_h^{{\scriptscriptstyle\perp}2}\!
\!-\!s_h^2 } {1\!+\!\bu_h^{{\scriptscriptstyle\perp}2}\!\!+\!s_h^2 }.  \label{u_hs_h}
\ea
\ee
The Cauchy problem (\ref{lageul}) with initial condition \ $\bx_{h}(0,\bX)\!=\!\bX$ \
is equivalent to the integral equations
\be
\ba{l}
\Delta z_e(x^0,Z):=z_h(x^0\!,Z)\!-\! Z \!=\!\! \displaystyle\int\limits^{x^0}_{0}\!\!
d\eta\, \beta_h^z[\eta,\! z_h(\eta,\!Z)],\\[8pt]
\bx^{{\scriptscriptstyle \perp}}_{h}(x^0\!,\!\bX )\!-\!\bX^{{\scriptscriptstyle \perp}}
\!=\!\! \displaystyle\int\limits^{x^0}_{0}\!\!
d\eta\, \bb_h^{{\scriptscriptstyle \perp}}[\eta,\! z_h(\eta,\!Z)]
\ea\qquad \label{inteq0}
\ee

Summarizing,  making use of  (\ref{hom'12}), (\ref{n_h}), (\ref{expl}),
(\ref{u_hs_h})  the evolution of the system is determined by
solving the system of integral equations (\ref{inteq1}),
 (\ref{inteq2}), (\ref{inteq0})$_1$ in the unknowns $\bA\!^{{\scriptscriptstyle\perp}}, s_h, z_h$
[note that, once this is solved, (\ref{inteq0})$_2$ becomes known].
It is natural to try an iterative resolution of the system 
(\ref{inteq1})+(\ref{inteq2})+(\ref{inteq0})$_1$ within the general  approach of the fixed point 
theorem:  replacing 
the approximation after $k$ steps [which we will distinguish by the superscript $(k)$]
at the right-hand side (rhs) of these equations we will
obtain at the left-hand side (lhs) the approximation after $k\!+\!1$ steps \cite{Fioetal}. 
If we are interested in solving the system for a short time interval after the beginning of
the interaction between the EM waves and the plasma,
and/or the initial densities are not very high, a convenient starting (0-th) step is the
zero-density solution $(\bu^{{\scriptscriptstyle\perp}}_e,
\tilde s_e,z_e)=(\bu^{{\scriptscriptstyle\perp(0)}}_e,1,z_e^{{\scriptscriptstyle(0)}})$. 
In next section we sketch the next approximation under some simplifying assumptions.

\section{Short pulse against a step-density plasma: the slingshot effect}
\label{constdensity}

\begin{figure}[ht]
%\begin{center}
\includegraphics[width=4cm]{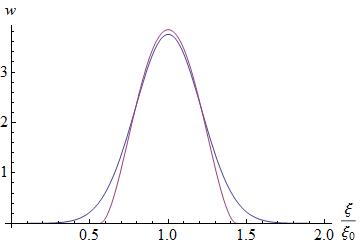}\hfill\includegraphics[width=3.6cm]{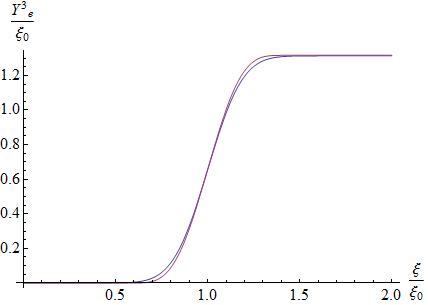}\hfill \includegraphics[width=4cm]{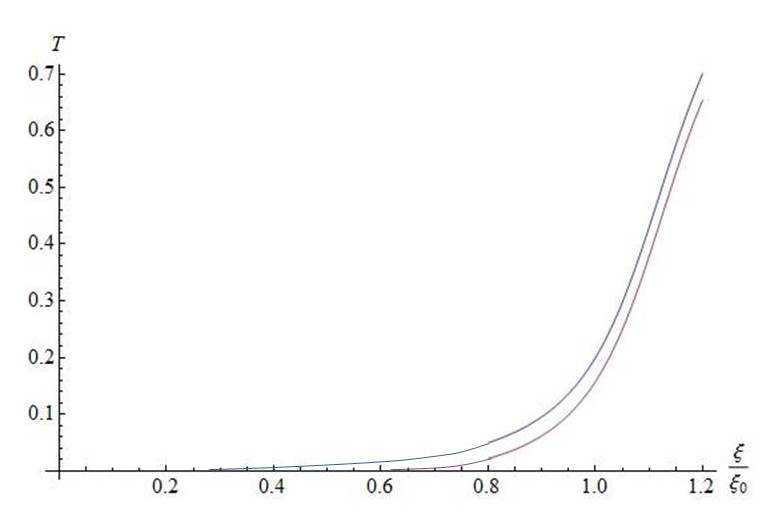}
%\end{center}
%\caption{The graphs of $w,Y^3_e,T$ for $w(\xi)\!=\!w_g(\xi)$ (blue line) and  $w(\xi)\!=\!w_p(\xi)$ and %$K\xi_0^2\!=\!2$, corresponding to $n_0\!\simeq\! 10^{18}cm^{-3}$. }
\label{Plots}       
\end{figure}

Henceforth we stick  to such small $x^0$ (small times after the beginning of the interaction)
that the motion of ions can be neglected (ions respond much more slowly
than  electrons because of their much larger mass). 
We formalize this by considering ions as infinitely massive, so that they remain 
at rest [$Z_h(x^0\!,\! z)\!\equiv\! z$ for $h\!\neq\! e$], 
have constant densities, and their contribution  to rhs(\ref{Maxwell12})
 disappears; only  electrons  contribute:
rhs(\ref{Maxwell12})$=\! 2\pi e n_e\bb^{{\scriptscriptstyle\perp}}_e$. 
Moreover we assume that
$\widetilde{n_{h0}}( Z )$ are not only zero for $Z\!<\!0$ but also constant for $Z\!>\!0$:
 $\widetilde{n_{e0}}( Z )\!=\!n_0\theta(Z)$, etc. (as depicted in fig. \ref{Plots}-left), where
$n_0$ is the initial electron and proton density.
Choosing $\bar Z\!=\!0$ \  in (\ref{expl}) we find
\bea
 E^{{\scriptscriptstyle z}}\!(x^0\!,z)\!=\!4\pi \sum\limits_{h=1}^k\! q_h
\widetilde{N}_h[ Z_h(x^0\!,z)]\!=\! 4\pi e n_0\left\{
z\,\theta(z)\!-\! Z_e(x^0\!, z)\,\theta[ Z_e(x^0\!, z)]\right\}\!\!.  \qquad       \label{elField}
\eea
If $z,Z\!>\!0$ this reduces to the known result (see e.g. \cite{Daw51,AkhPol67}) 
that at $x^0$ the electric force acting on the electrons initally located
in $Z$ is of the harmonic type 
\ $\widetilde{F}_e^{{\scriptscriptstyle z}}(x^0\!,Z)\!=\!- 4\pi n_0 e^2
\Delta z_e(x^0\!, Z)$, \ i.e. proportional to their displacement \ 
$\Delta z_e(x^0\!, Z)\!:=\! z_e(x^0\!, Z)\!-\! Z$ \ w.r.t. their initial position.

The corresponding first corrected approximation reads \cite{Fio13}:
\bea
&& \bu_e^{{\scriptscriptstyle\perp(1)}}(x^0\!,z)-\bu_e^{{\scriptscriptstyle\perp(0)}}(x^0\!\!-\!z)
=-\frac{2\pi e^2}{m_ec^2} \!\!\displaystyle\int_{D\!^0_{x}}\!\!\!\! d^2x'\:
\left[n_e^{{\scriptscriptstyle(0)}}\bb^{{\scriptscriptstyle\perp(0)}}_e \right]\!\!(x'),        \nn 
&& \tilde s_e^{{\scriptscriptstyle(1)}} =e^{\tilde r_e^{{\scriptscriptstyle(0)}}}, \label{appr1}\\
&& \Delta z_e^{{\scriptscriptstyle(1)}}(x^0,Z)= \!\! \displaystyle\int\limits^{x^0}_{0}\!\!
d\eta\,\tilde \beta_e^{{\scriptscriptstyle z(1)}}(\eta, Z ).\nonumber
\eea
Here the change from the Eulerian to the  Lagrangian description (represented by the tilde)
is performed approximating $\bx_e$ by $\bx_e^{{\scriptscriptstyle(0)}}$; in the second line 
the integral corresponding to the second term of (\ref{inteq2}) does not appear because
$ \partial\!_{{\scriptscriptstyle -}}\bu_e^{{\scriptscriptstyle\perp(0)}2}\!=\!0$, and we have 
abbreviated
\bea
\tilde r_e^{{\scriptscriptstyle(0)}}(x^0\!,\! Z ):=4K\!\!\int^{x^0}_{Z}\!\!d\eta\frac {z_e^{{\scriptscriptstyle(0)}}\theta\big[z_e^{{\scriptscriptstyle(0)}}\big]\!-\! Z\theta(Z)}
{\tilde\gamma_e^{{\scriptscriptstyle(0)}}}(\eta,\! Z ) =
4K\,V_e^3\!\left[\Xi_e^{-1}\!( x^0\!\!-\! Z)\right],\nn
\mbox{where }\qquad K\!:=\!\frac{\pi e^2 n_0}{m_ec^2},\qquad  V_e^3(\xi):=\!
\displaystyle\int\limits^{\xi}_0\!\! dy\, Y^3_e(y)\quad [Y^3_e\mbox{ defined in } (\ref{defYXi})].       \nonumber  %\label{defr}
\eea
If the EM wave is  (\ref{prototype}) with \ $\lambda |\epsilon_s'/\epsilon_s|\!\le\!\delta\!\ll\!1$,  \ 
setting \  $w\!:=\!\frac{ e}{km_ec^2}\epsilon_s$   \ one finds
$$
\Ba_e^{{\scriptscriptstyle \perp}}\!\simeq\! \frac{1}{k}\epsilon_s\Be_p^{{\scriptscriptstyle \perp}}\!,\qquad
\bu_e^{{\scriptscriptstyle \perp(0)}}\!\simeq\! w\Be_p^{{\scriptscriptstyle \perp}}\!,\qquad
\bY_e\!^{{\scriptscriptstyle \perp}}\!\!\simeq\! -\frac 1k w\Be_o^{{\scriptscriptstyle \perp}}\!,\qquad
u_e^{{\scriptscriptstyle z(0)}}\!\simeq\! \frac 12 w^2\Be_p^{{\scriptscriptstyle \perp 2}},
$$
where \ $a\!\simeq\!b$ \ means \
$a\!=\!b\!+\!O(\delta)$.  From (\ref{appr1}) one can show \cite{Fio13} that 
%we can approximate 
\be
\bA_e^{{\scriptscriptstyle \perp(1)}}\!\simeq\!\Ba_e^{{\scriptscriptstyle \perp}},\quad
\bu_e^{{\scriptscriptstyle \perp(1)}}\!\simeq\!\bu_e^{{\scriptscriptstyle \perp(0)}} \qquad\:  
\mbox{if }\quad\: 0\!\le\!x^0\!\!-\!z\!\le\! \xi_0,\quad
0\!\le\!x^0\!+\!z\!\ll\!  \frac{2\pi} {K\lambda}.    \label{stregion}
\ee
($\xi_0$ stands for the first maximum point of $w, \epsilon_s$) 
by showing that the relative difference between the lhs and the rhs is much smaller than 1 in the
spacetime region (\ref{stregion})$_2$. There  we find in particular,
by (\ref{displace}), (\ref{u_hs_h}) and some computation,
\bea
&& \beta_e^{{\scriptscriptstyle z(1)}}\!=\!\frac {1\!+\!\bu_e^{{\scriptscriptstyle\perp(1)}2}
\!\!-\!s^{{\scriptscriptstyle (1)}2}_e}{1\!+\!\bu_e^{{\scriptscriptstyle\perp(1)}2}\!\!+
\!s^{{\scriptscriptstyle (1)}2}_e }
\!\simeq\!\frac {1\!+\!\bu_e^{{\scriptscriptstyle\perp(0)}2}
\!\!-\!e^{2r^{{\scriptscriptstyle (0)}}_e}}{1\!+\!\bu_e^{{\scriptscriptstyle\perp(0)}2}\!\!+
\!e^{2r^{{\scriptscriptstyle (0)}}_e} },                      \qquad   \label{beta1}\\
&&\Delta z_e^{{\scriptscriptstyle (1)}}(x^0\!,\! Z )\simeq 
% S[\Xi_e^{-1}\!( x^0\!\!-\! Z)],\qquad S(\xi)\!:=\!\!\int^{\xi}_0\! dy\, 
%[\gamma_e^{{\scriptscriptstyle (0)}}\beta_e^{{\scriptscriptstyle z(1)}}](y)\qquad
\!\!\!\!\!\!\!\displaystyle\int\limits^{\Xi_e^{-1}\!( x^0\!\!-\! Z)}_0
\!\!\!\!\!\!\!\! dy\, [\gamma_e^{{\scriptscriptstyle (0)}}\beta_e^{{\scriptscriptstyle z(1)}}](y)
%\!\gtrsim\!\Sigma\!\left[\Xi_e^{-1}\!( x^0\!\!-\! Z)\right]
,\qquad\\
&& \ba{l} 0\le [\Delta z_e^{{\scriptscriptstyle (0)}}\!-\!\Delta z_e^{{\scriptscriptstyle (1)}}](x^0\!,\! Z )
\simeq G[\Xi_e^{-1}\!( x^0\!\!-\! Z)], \\[6pt]
G(\xi)\!:=\!\!\int\limits^{\xi}_0\! dy\, g(y), \quad
g:=\frac{\left(\!1\!+\!2u_e^{{\scriptscriptstyle z(0)}}\!\right)\! 
\big(e^{2r^{{\scriptscriptstyle (0)}}_e}\!\!\!-\!1\!\big) }
{1\!+\!2u_e^{{\scriptscriptstyle z(0)}}\!+\!e^{2r^{{\scriptscriptstyle (0)}}_e}},
\ea\qquad\\[8pt]
&& 0\le \frac{\Delta z_e^{{\scriptscriptstyle (0)}}\!-\!\Delta z_e^{{\scriptscriptstyle (1)}}}
{\Delta z_e^{{\scriptscriptstyle (0)}}}(x^0\!,\! Z )
\simeq T[\Xi_e^{-1}\!( x^0\!\!-\! Z)],\qquad\qquad
T\!:=\!\frac{G}{Y_e^3}.\qquad\label{reldif}
\eea{}
The last expression is the relative difference 
between the displacement $\Delta z_e$ in the zero-density and in the first corrected approximation. 
Hence the approximation \ $z_e(x^0\!,\! Z )\!\simeq\! z_e^{{\scriptscriptstyle(1)}}(x^0\!,\! Z )\!\simeq\! z_e^{{\scriptscriptstyle(0)}}(x^0\!,\! Z )$
may be good only as long as \ $T[\Xi_e^{-1}\!( x^0\!\!-\! Z)]\!\ll\! 1$. By (\ref{chain}),  the maximum $\Ba^{{\scriptscriptstyle \perp}}(\xi_0)$ reaches  the electrons initially located in $Z$ at the time $\check x^0(Z)\!=\!\Xi_e(\xi_0)\!+\! Z$;  
therefore the approximation \ $z_e(x^0\!,\! Z )\!\simeq z_e^{{\scriptscriptstyle(0)}}(x^0\!,\! Z )$
may be good for all $x^0\!\le\!\check x^0(Z)$ only if
\be
%T[\Xi_e^{-1}\!( \check x^0\!-\! Z)]\!=\!
T(\xi)\ll 1\qquad\quad 0\le\xi\le\xi_0,
\qquad\qquad\quad
%\check x^0\!+\!z_e^{{\scriptscriptstyle (0)}} (\check x^0\!, Z )\!=\!
2Y^3_e(\xi_0)\!+\!\xi_0\!+\! 2Z \ll  \frac{2\pi} {K\lambda}.          \label{condgood}
\ee
In particular,  (\ref{checkzeta}) with \ $\check\xi\!=\!\xi_0$
will give a good estimate $\zeta_e$ of the displacement of the plasma-surface electrons
(those with $Z$ close to zero) if (\ref{condgood}) is satisfied.

In \cite{FioFedDeA13} $\zeta_e$ is used to predict
and estimate the  {\it slingshot effect}, i.e. the expulsion of 
very energetic electrons in the negative $z$-direction  shortly after the impact of
a suitable ultra-short and ultra-intense laser pulse in the form of a {\it pancake}
(i.e. a cylinder of radius $R$ and height $l\!\ll\!R$) 
normally onto a plasma.  The mechanism is very simple: the plasma electrons in 
a thin layer - just beyond the surface of the plasma - first are given sufficient 
electric potential energy by the displacement  $\zeta_e$  w.r.t. the ions, 
then after the pulse are pulled back by the longitudinal 
electric force exerted by the latter and may leave the plasma. 
Sufficient conditions for this to happen are: 1.  $l\!\ll\!R$,
so that plane wave solutions are sufficiently accurate within the plasma, 
especially in the forward boost phase;
2. $R\!\gtrsim\!2\zeta$, to avoid trapping of the boosted electrons or 
even the onset of the {\it bubble regime} \cite{MalEtAl05}; 
3. the EM field inside the pancake is sufficiently intense, and/or
$n_0$ is sufficiently low, so that the longitudinal electric force induces
the  back-acceleration of the electrons mainly
{\it after}  the pulse maximum has overcome them (in phase with the negative
ponderomotive force exerted by the pulse in its decreasing stage).
Actually we impose the stronger condition that $n_0$ is sufficiently small in order that 
(\ref{condgood}) be fulfilled and the estimate $\zeta_e$ be reliable. 
As a result, an estimate of the final  energy of the electrons initially 
located at  $Z\!=\!0$ after the expulsion is  \cite{FioFedDeA13}
\be{}
H=m{}c^2\gamma_{e{\scriptscriptstyle M}},\qquad\qquad
\gamma_{e{\scriptscriptstyle M}}\simeq  1+2K \zeta_e^2.
%\simeq 1+2K \,\left[Y_e^3(\xi_0)\right]^2.       
\label{gammaeM}
\ee
The above conditions are already at hand in several laboratories. 
The resulting $H$ would be of few  MeV implementing those 
available at the FLAME facility (LNF, Frascati), or at the ILIL laboratory (CNR, Pisa):
the pulse energy is a few joules, \
$\lambda\sim  10^{-4}cm$, \ $\xi_0\sim 10^{-3}cm$, \ $K\xi_0^2\!\sim\!1$ 
(whence $n_0\!\sim\! 10^{18}cm^{-3}$);
(\ref{condgood}) are fulfilled, see the typical plots  reported below
[the blue, purple curves resp. correspond to a gaussian and to a cut-off
polynomial amplitude $w(\xi)$].

%\begin{acknowledgements}
%If you'd like to thank anyone, place your comments here
%and remove the percent signs.
%\end{acknowledgements}

% BibTeX users please use one of
%\bibliographystyle{spbasic}      % basic style, author-year citations
%\bibliographystyle{spmpsci}      % mathematics and physical sciences
%\bibliographystyle{spphys}       % APS-like style for physics
%\bibliography{}   % name your BibTeX data base

% Non-BibTeX users please use

\end{document}